\documentstyle[12pt]{article}
\newlength{\dinwidth}
\newlength{\dinmargin}
\def\sel{\sigma_{el}^{VN}}
 \def\sin{\sigma_{in}^{VN}}
 \def\stot{\sigma_{tot}^{VN}}
 \def\inf{\int_{-\infty}^{\infty}}
 
 \newcommand\beq{\begin{equation}}
 \newcommand\eeq{\end{equation}}
 \newcommand\beqn{\begin{eqnarray}}
 \newcommand\eeqn{\end{eqnarray}}
 \newcommand{\doublespace} {
 \renewcommand{\baselinestretch} {1.6}
 \large\normalsize}

 \newcommand{\ra}{\rangle}

 \def\sel{\sigma_{el}^{VN}}
 \def\inf{\int_{-\infty}^{\infty}}
\begin{document}
\vspace*{1cm}

\begin{center}
{\Large {\bf Electroproduction of vector mesons off nuclei at ELFE~:\\
exact solution for color transparency}\footnote{
A talk presented by B.K. at the second ELFE Workshop,
September 22-27, 1996, Saint Malo, France}}\\

\vspace*{5mm}
{\large J\"org~H\"ufner$^a$ and 
Boris~Kopeliovich$^{bc}$\footnote{Partially 
supported by the European Network:
Hadronic Physics with Electromagnetic Probes. Contract 
No. FMRX-CT96-0008}}\\
\end{center}
$^a$ Inst. f\"ur Theor. Physik der Universit\"at ,
 Philosophenweg 19, 69120 Heidelberg, Germany\\
$^b$ Max-Planck Institut f\"ur
 Kernphysik, Postfach
103980,  69029 Heidelberg, Germany\\
$^c$ Joint Institute
 for Nuclear Research,
Dubna, 141980
 Moscow Region, Russia\\

\begin{abstract}
A multichannel evolution equation is developed 
for the density matrix
describing a hadronic wave packet
produced by a virtual photon
and propagation through a nuclear matter.
This approach is dual to the quark-gluon representation, 
incorporates the effects of coherence
and formation times and gives an exact solution
for color transparency problem at any photon energy.
We propose a procedure of data analysis, which
provides an unambiguous way of detection of a color
transparency signal even at medium energies.
\end{abstract}
\doublespace
\section{Introduction}
Exclusive electroproduction of vector mesons
was suggested in \cite{knnz} as an effective tool in 
search for color transparency (CT) \cite{prp}. 
The key idea is based upon
absence of strict correlation between the photon's 
energy and virtuality, typical for reactions of 
quasielastic scattering. Data from the E665 experiment
\cite{e665} nicely confirms the predicted value of 
the effect \cite{knnz}. The statistical confidence of the
observed growth of nuclear transparency with the photon
virtuality $Q^2$ is, however, quite modest and new experiments
are planned at lower energies (HERMES, TJNAF, ELFE).\\
Although CT phenomenon is most naturally interpreted in 
the quark-gluon representation,
the calculations are easy only at energies $\nu \gg Q^2R_A/2$,
when size of the photon fluctuations is frozen during
propagation through the nucleus. At lower energies the produced
colorless wave packet is developing while it propagates through
the nucleus. Accordingly, absorption in nuclear matter varies
and the expected CT effect may be substantially reduced.
Such an evolution is controlled by the so-called formation time $t_f$,

\beq
\frac{2\nu}{Q^2} < t_f < \frac{2\nu}{m_{V'}^2 - m_V^2}\ .
\label{1}
\eeq

The bottom limit
corresponds to most quickly expanding states of small size,
$r_T^2 \propto 1/Q^2$. In order to observe a full effect of
CT, one should make this time , $t^{min}_f \approx 1/m_Nx_B \gg R_A$,
where $x_B$ is the Bjorken variable. 
The upper limit in (\ref{1}) corresponds to a 
long evolution of a rather large-size
wave packet consisted mostly from the two lightest 
states, $V$ and $V'$. This time $t_f^{max}$ controls the onset of CT.\\
There is another phenomenon, which follows from 
the quantum-mechanical uncertainty for the production time of 
the final wave packet, which is usually called coherence time,

\beq
t_c \approx \frac{2\nu}{Q^2+m_V^2}
\label{2}\ .
\eeq

This uncertainty may be interpreted as a lifetime of
hadronic fluctuations of the photon. If this time
substantially exceeds the nuclear size, $t_c \gg R_A$,
one deals with a virtual 
hadronic, rather than with photonic beam. Correspondingly,
nuclear attenuation increases \cite{hkn2,hkn1,kn1}. \\
Two methods for correct calculation of the wave packet evolution
are known. One was developed in \cite{kz} 
using the quark-gluon representation and Feynman path integral technique.
However, as far as the coherence time effects are involved, the 
path integrals approach becomes unreasonably complicate, and no 
solution is still found.\\
Another approach uses the hadronic basis for 
the photon fluctuations, which is dual
to the quark-qluon basis, provided that completeness takes place.
An exact solution, incorporating 
both coherence and formation time effects 
is found and the results are presented in this talk. 
\section{Coherence time}
Coherence (interference) of the vector meson waves produced
at different longitudinal coordinates is important both for
coherent (the nucleus remains intact) and incoherent (the nucleus
breaks up) electroproduction.
Effect of coherence time exists even in Glauber
approximation, although a correct formula for incoherent
electroproduction was derived only recently \cite{hkn2}.
 Vector mesons produced at different points
 separated by longitudinal
distance $\Delta z$ have
 a relative phase shift $q_c\Delta z$, where 
$q_c = (Q^2 + m_V^2)/2\nu$ is the
longitudinal momentum transfer in 
$\gamma^*N \to VN$, 
$Q^2$ and $\nu$ are 
the virtuality and energy
 of the photon, respectively.
Taking this into account one arrives at the following
expression \cite{hkn2} for nuclear transparency defined as 
$Tr=\sigma_A/A\sigma_N$,

\beqn
 & &Tr_{inc} =
\frac{\stot}{2A\sel}(\sin-\sel)
 \int d^2b\ \inf dz_2\ \rho(b,z_2)
\int_{-\infty}^{z_2} dz_1\ \rho(b,z_1)\
 \nonumber\\
 &\times &
e^{iq_c(z_2-z_1)}\
 \exp\left[ -{1\over 2}\stot\ \int_{z_1}^{z_2}dz
\rho(b,z)\right]
 \exp\left[ -\sin\ \int_{z_2}^{\infty} dz\
\rho(b,z)\right]\
 \nonumber\\
 &+&
 \frac{1}{A\sin}
 \int d^2b \left
[1-e^{-\sin T(b)}\right ]
 -Tr_{coh}\ ,
\label{3}
\eeqn

 \beq
Tr_{coh} = \frac{(\stot)^2}{4A\sel}
 \int d^2b\left |\inf dz\
 \rho(b,z)\
e^{iq_cz}
 \exp\left[-{1\over 2}\stot\int_z^{\infty}
 dz'\rho(b,z')\right
]\right |^2\ ,
\label{4}
\eeq
 
where $Tr_{coh}$ 
corresponds to the coherent case \cite{bauer,hkn2,hkn1}.
$Tr_{inc}$, in contrast to the coherent case \cite{kn1},
decreases with energy from $Tr_{inc}=
 \sigma_{in}^{VA}/A\sigma_{in}^{VN}$ 
($q_c \gg 1/R_A$) down to
$Tr_{inc}=\sigma_{qel}^{VA}/A\sigma_{el}^{VN}$
($q_c \ll 1/R_A$).
Numerical examples are presented in \cite{hkn1,kn1,hkn2}.\\
Variation of $t_c$ 
 may be caused either by its $\nu$- 
or $Q^2$-dependence. In the latter case $t_c$ decreases
with $Q^2$ and the nuclear transparency 
grows, what is usually expected to be a signature of CT
\cite{knnz}.
Our results for incoherent
electroproduction of $\rho$-meson on lead are 
shown by dashed curves in Fig.~1 (more examples are 
in \cite{hkn1,kn1,hkn2}). The predicted  $Q^2$-dependence
is so steep that makes it quite problematic to observe a
signal of CT on such a background.
\section{Formation time}
Inclusion of excited states of the vector meson into the multiple
scattering series is known as Gribov's inelastic corrections \cite{gribov}.
CT corresponds to a special tuning of these corrections, when
the diagonal and off diagonal amplitudes cancel in
final state interaction at high $Q^2$. The amplitudes
we use satisfy such a condition,
since we calculate the photoproduction
amplitudes projecting to the $Q^2$-dependent 
$q\bar q$ component of the 
photon wave function.\\
In the case of incoherent production one has to sum over
all final states of the nucleus. Therefore, the
the wave packet should be 
described by density matrix $P_{ij} = 
\sum_{A^*}|\psi_i\ra|\psi_j\ra^+$.
Wave function $|\psi_i\ra$,
 has components
$\gamma^*$, $V$, $V'$, etc.
The evolution equation for the density matrix reads \cite{hk1},

 \beq
i\frac{d}{dz}\widehat P =
 \widehat Q \widehat P -
 \widehat P \widehat Q^+ -
 {i\over 2}\sigma^{V N}_{tot}
 \left(\widehat T \widehat P +
 \widehat
P \widehat T^+\right) +
 i\sigma^{VN}_{el}
 \widehat T \widehat P
\widehat T^+\ , 
\label{11}
\eeq

 \beq
 \widehat Q =
 \left(\begin{array}
{cccc}0&0&0&...\\0&q&0&...\\0&0&q'&...
\\.&.&.&...
\end{array}\right)\ ;\ \ \ \ 
 \widehat T =
 \left(\begin{array}
 {cccc}0&0&0&...\\\lambda&1&\epsilon&...
 \\\lambda R&\epsilon&r&...\\
.&.&.&...
\end{array}\right)\ ,
\label{8}
\eeq

where $\widehat Q$ and $\widehat T$ are the $(n+1)\times(n+1)$ 
matrices, and $n$ is the number
of states involved into consideration. 
$q,\ q'\ ...$ 
are the transferred longitudinal momenta,
 $q(q') = (m^2_{V(V')} + Q^2)/2\nu$. 
For other parameters we use notations from
\cite{hk}, $r=\sigma_{tot}^{V'N}/\sigma_{tot}^{VN}$,
$\epsilon = f(VN \to V'N)/f(VN \to VN)$ and
$R= f(\gamma N \to V'N)/f(\gamma N \to VN)$.
The value of parameter $\lambda = 
f(\gamma N \to V'N)/f(VN \to VN)$
is inessential, since it cancels in nuclear transparency.
The boundary condition for the density matrix is
$P_{ij}(z\to -\infty) = \delta_{i0}\delta_{j0}$.
Note, eq.~(\ref{11}) reproduces 
(\ref{3}) if $\epsilon=0$ and eq.~(\ref{4}) 
if $\sigma_{el}^{VN} = 0$.\\
We calculated the $Q^2$-dependence of nuclear transparency for incoherent
electroproduction of $\rho$-meson on lead
in two-channel approximation, using the parameters in (\ref{8})
as in \cite{kn1} (more examples, including coherent production,
radial excitations and other flavours are in \cite{kn1,hk1}).
The two channels should well reproduce
the onset of CT, while at higher energies, $1/x_B \ll m_NR_A$
the full CT may develop only after inclusion the the higher excitation states.
The results for different energies are shown by solid curves
in Fig.~1.
\begin{figure}[tbh]
 \includegraphics{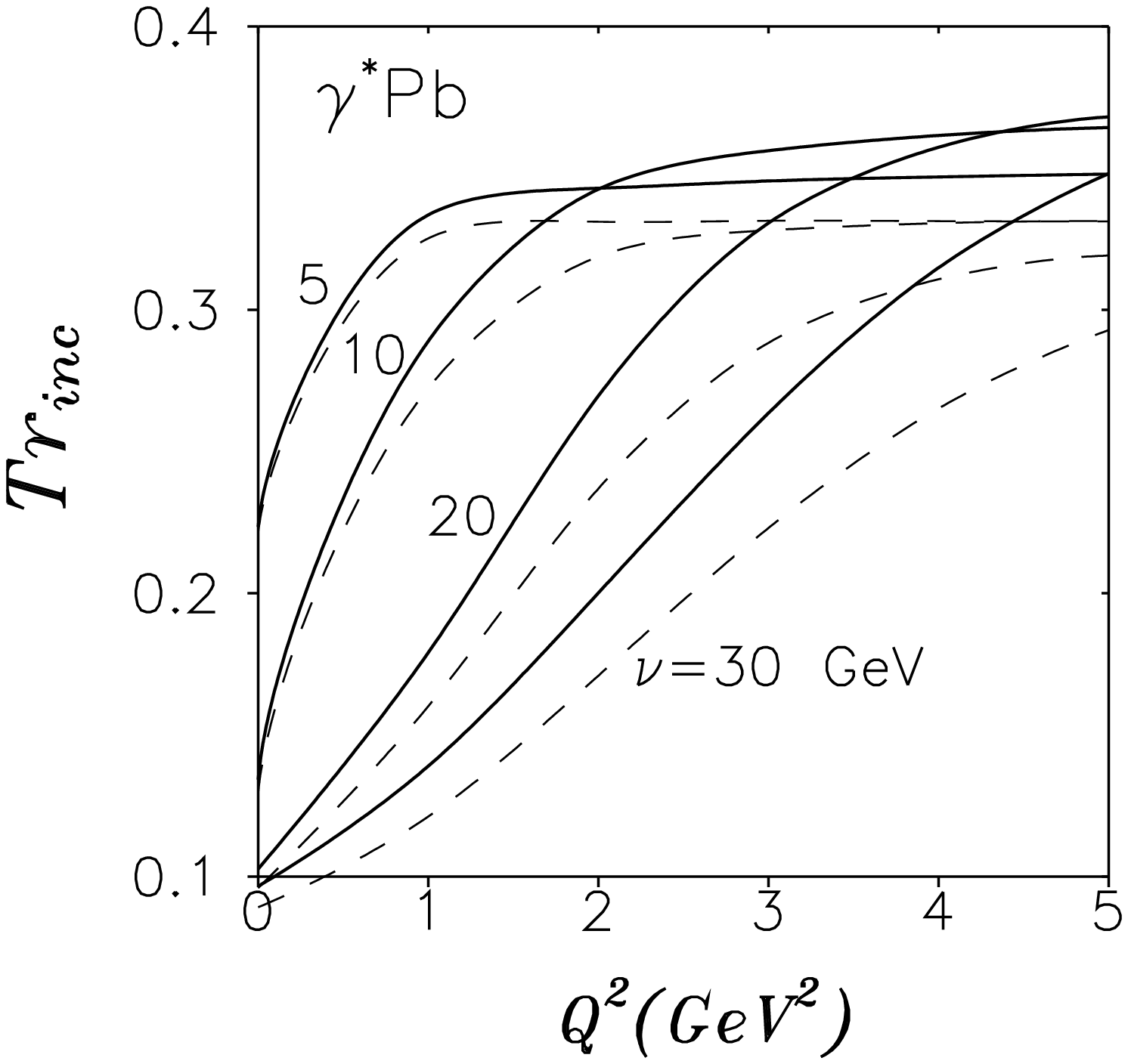}
\includegraphics{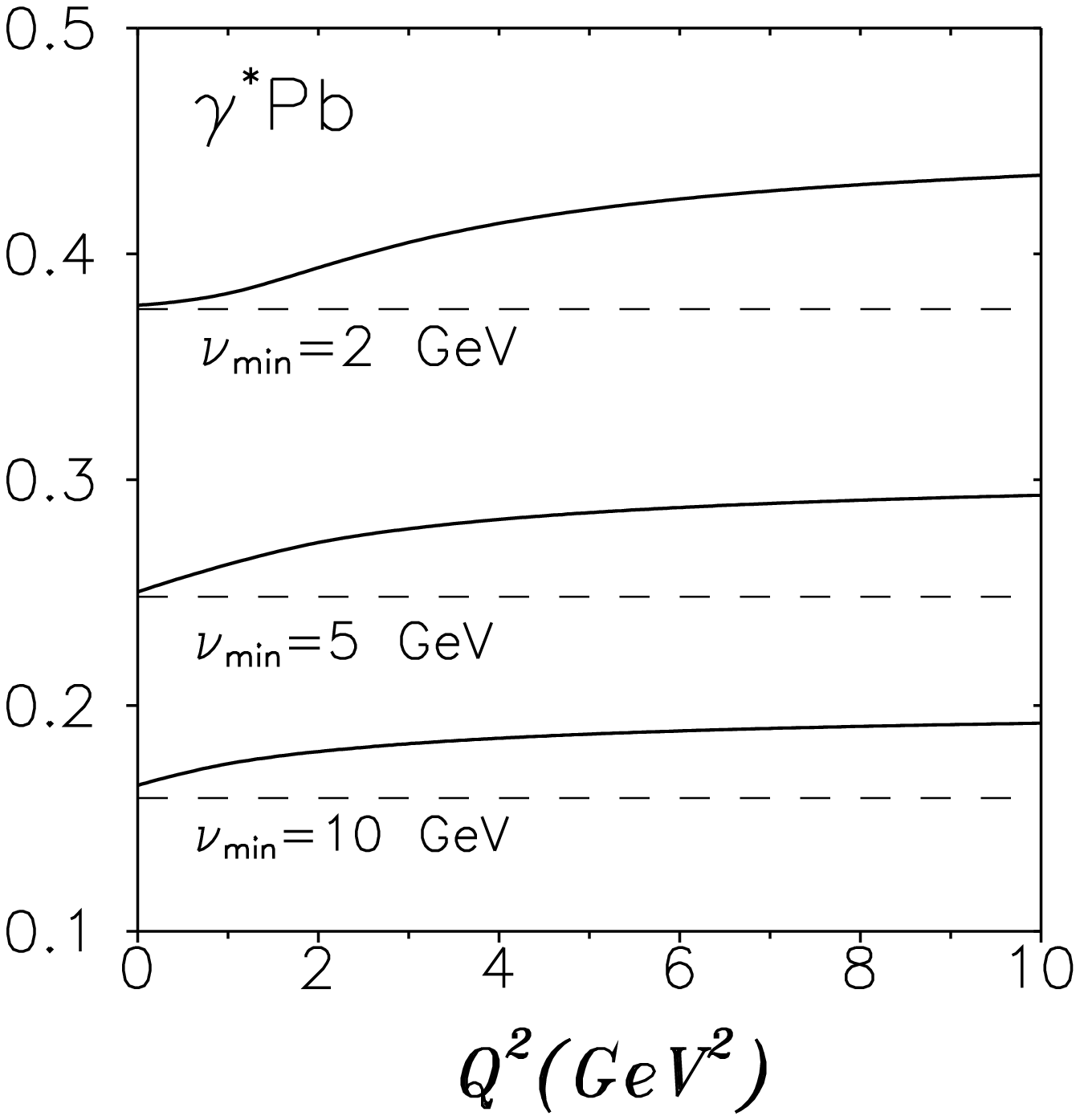}
\vspace{6.5cm}
\parbox{14cm}
 {\caption[Delta]
 {\it $Q^2$-dependence of nuclear transparency for
$\rho$-meson electroproduction on lead at
photon energies $\nu = 5,\ 10, 20$ and $30\ GeV$.
Dashed curves correspond to Glauber approximation,
solid curves are calculated with the evolution 
equation eq.~(\ref{11}).}
\label{fig1}}
\parbox{14cm}
{\caption[Delta]
 {\it The same as in Fig.~1, but with fixed 
$t_c=2\nu_{min}/m_{\rho}^2$.}
\label{fig2}}
\end{figure}
Although the growth of nuclear transparency is steeper than what we expect
in Glauber approximation, the difference is too small to be used as a signature
of CT. Even the present state of art of Glauber-model calculations
leaves enough freedom to fit in such a narrow corridor in nuclear 
transparency.\\
At this point we would like to soften our pessimism and 
suggest a method for unambiguous detection of onset of CT.
The key idea is quite straitforward \cite{hk1}: as soon as
the variation of the coherence time may mock the CT effects,
one should fix $t_c$. This can by done by means of 
a special selection
of events with different $Q^2$ and $\nu$.
Starting from minimal energy
$\nu_{min} = t_cQ^2/2$ with real photoproduction one
should increase both $Q^2$ and $\nu$, while $t_c = const$,
in accordance with
(\ref{2}). Our predictions for $Q^2$-dependence of the lead 
transparency at different values of minimal energy (or $t_c$)
are depicted in Fig.~2 in comparison with $Q^2$-independent
expectations of Glauber approximation.

Concluding,
we have developed a multichannel
approach to incoherent exclusive electroproduction of
vector mesons off nuclei, which incorporates the effects of
coherent and formation times, as well as CT. 
Variation of the coherence time with the photon energy and $Q^2$ 
causes substantial changes of the nuclear transparency and may mock
onset of CT. We suggest such a mapping 
of $\nu$ and $Q^2$ values, which keeps the coherence time constant.
This helps to single out an unambiguous signal of CT at medium energies.

\end{document}